\documentclass[aps,twocolumn]{revtex4}
 
\usepackage{epsfig}       

\begin{document}  

\title{A minimal mechanism leading to discontinuous phase transitions
for short-range systems with absorbing states}               
\author{Carlos E. Fiore} 
\affiliation{Departamento de F\'{\i}sica, Universidade Federal do Paran\'a \\
Caixa Postal 19044, 81531-000 Curitiba, Paran\'a, Brazil}

\date{\today}

\begin{abstract}
Motivated by recent findings, we discuss the existence of a direct and
robust mechanism providing  discontinuous absorbing
transitions in short range systems with single species, with no extra
symmetries or conservation laws.
We consider  variants of the contact process, in 
which at least two adjacent particles (instead of one, as commonly 
assumed) are required to create a new species. 
Many interaction rules are analyzed, including distinct cluster annihilations 
and a modified version of the original pair contact process (PCP).
Through detailed time dependent numerical simulations we find that
for our modified models, the phase transitions are of first-order,
hence contrasting with their corresponding usual formulations in the
literature, which are of second-order.
By calculating the order-parameter distributions,
the obtained bimodal shapes as well as the finite scale analysis
reinforce coexisting phases, so a discontinuous transition.
These findings  strongly 
suggest that  above particle creation requirements 
constitute a minimum and fundamental mechanism 
determining the phase coexistence in short-range contact processes.

\end{abstract}

\maketitle

{\it Introduction.} Nonequilibrium  phase transitions into absorbing states  
have attracted great interest in recent years, 
not only for the possibility of describing a countless number of processes,
such as    wetting phenomena, spreading of diseases, 
chemical reactions and others \cite{marro,odor04}
but also for the searching of  experimental realizations \cite{sano}.
In the simplest examples, they 
manifest in single species systems, such as
probabilistic cellular automata or contact processes (CP) \cite{marro,harris}. 
Typically,  these transitions are second-order belonging to 
 the directed percolation (DP) universality class \cite{odor04}.
Although few frequent in above situations, discontinuous 
absorbing transitions have also been observed.  
Mean-field  approaches \cite{schogl},   lattice models  
\cite{zgb,lubeck,toom} or continuous 
descriptions \cite{janssen} reveal that  its occurrence 
requires an effective mechanism that suppresses low density 
states.  
According to the  Elgart and Kamenev classification \cite{elgart},   for
one-component reaction diffusion  with $n-$particle creation
and $k-$particle annihilation, the reactions
$kA\rightarrow (k-l)A$ and  $nA \rightarrow (n+m)A$ summarize
the existence of a discontinuous transition whenever  $k<n$. 
Although such semi-classical field theory is an important  
benchmark, suggesting  
crucial ingredients for its occurrence, the system
dimensionality or the inclusion of
spatial fluctuations  may
suppress the stabilization of compact clusters 
in the above conditions \cite{hinri00,park08}.

\begin{figure}
\centering
\includegraphics[scale=0.28]{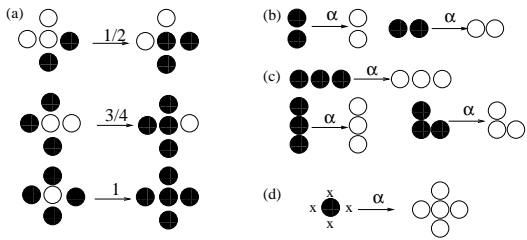} 
\caption{Some  examples of transition rates in a square
lattice ($z=4$). The models A, B and C 
are defined by interaction rules $(a)-(b)$, 
$(a)-(c)$ and $(a)-(d)$, respectively. In $(d)$, the symbols
$\times$  denote a  local configuration composed of $nn$  nearest
neighbor occupied sites (with $nn$ ranged from 0 to 4) 
and after the annihilation all $nn+1$ particles are extinct.}
\label{fig0}
\end{figure}

Inspired by mean-field like predictions \cite{schogl}, some {\it
  restrictive} versions of the two and three dimensional contact
process (CP) \cite{harris} have been considered
\cite{evans,evans2,oliveira}. They differ from the original case 
in which more than one nearest neighbor occupied sites are required to create a new
particle  (instead of one as in the standard CP), and single particles
are annihilated. 
In the simplest
case \cite{oliveira}, two particles are required and  the 
creation does not depend on the
specific particle displacements, as exemplified in Fig. \ref{fig0} $(a)$.  
Unlike the original CP, the
transition becomes discontinuous for dimensions larger than $1$.
Extension of such interactions for complex networks \cite{durret} (instead of regular
lattices \cite{evans,evans2,oliveira}) have revealed that the topology
of the lattice does not affect the phase coexistence.  On the other
hand, by changing the dynamics mildly, where one nearest neighbor pair
is necessary to create a new offspring ({\it instead of two nearest
  neighbor particles but still fulfilling the condition $k<n$}) the
phase coexistence is suppressed, returning to be continuous
(schematically, such change is equivalent to shift the local rule 
of particle creation at $0$ from
$1-0-1$ to $1-1-0$).  All these comments inspire us to raise two
fundamental questions: Is there an ingredient that always provides a
discontinuous absorbing transition in single species systems? If so,
what is this dynamics?  To try to answer 
such questions, we investigated thoroughly a class of four restrictive
processes. In the first three examples, we consider the particle
creation in the presence of at least two particles,
as considered in Ref. \cite{oliveira} and a 
family of annihilation processes (to be described further). Our goal
is to verify if the phase coexistence is mantained
by changing the annihilation rules.   
The fourth model is a small modification in
the pair contact process (PCP), a notorious model with infinitely many
absorbing states and a DP phase transition \cite{jensen,dic99}. 
In our modified version, at least two pairs of particles 
are required (instead of one as in the original PCP) for creating a new
particle. This  modification aims to verify if, in similarity with
previous cases, this small change is sufficient for
shifting the order of transition.
As will be shown, under two distinct methodologies,  
in all restrictive models the phase transition is first-order,
what suggests that the particle creation
in the presence of a minimal neighborhood (for the studied models 
it is 2) 
constitute a  fundamental (and robust) 
mechanism ruling   discontinuous absorbing phase transitions.

{\it Models and methods.} 
In all the situations considered here, if a site $i$ is empty 
(occupied) then the occupation variable $\eta_i$ assumes the value 0 (1).
In the first three model versions, when $\eta_i = 0$ a particle
can be created at $i$ with a probability $nn/z$ for $nn \geq 2$ and
zero otherwise (Fig. \ref{fig0} (a)). There is no creation
if $nn \le 1$.
Here, $nn$ is the number of occupied neighbors of $i$ and $z$
is the lattice coordination number. In a square lattice (all our studies)
$z$ reads $4$ .
In a similar fashion, particles can be annihilated with a probability
$\alpha$ (according to the rules described below).

\begin{figure}
\centering
\includegraphics[scale=0.28]{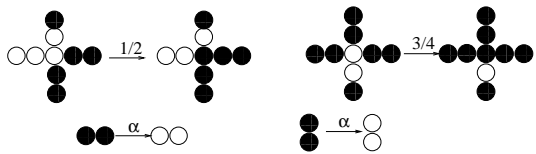} 
\caption{Some  examples of transition rates for the model D. Note
that there  is no particle creation if the number of pairs
of particles $nn_p$ is smaller than 2.}
\label{fig00}
\end{figure}

In the model A, annihilation occurs only for pairs of adjacent particles 
($k=l=2$).
So, an isolated particle cannot be destroyed, Fig. \ref{fig0} $(b)$.
For model B, annihilation occurs only for three adjacent particles 
($k=l=3$),
Fig. \ref{fig0} $(c)$. So, neither isolated nor pairs
of particles are eliminated.
Finally, in model C the annihilation of a particle at site $i$ 
automatically wipes out all its $nn$ nearest neighbors occupied
sites ($k=l=nn+1$),  Fig. \ref{fig0} $(d)$. 
Therefore, contrasting with A and B, the number of exterminated particles 
is not fixed, ranging from $1$ to $5$ in a square lattice 
(recall $nn$ varies from $0$ to $4$ in a square lattice). 
Thus, for the models B and C, the Elgart and Kamenev
classification is violated.
The last one, model D, is a modification of the PCP \cite{jensen,dic99}.
In the original PCP, only pairs of particles are annihilated and a 
particle is created with probability $nn_{p}/z$ if the number of 
neighboring pairs $nn_{p} \geq 1$.
In the model D we consider that the creation can occur only if 
$nn_{p} \geq 2$ (Fig. \ref{fig00}). 

For any of the above model versions, a phase transition is expected to 
separate an active regime (stable for low $\alpha$) from an absorbing 
phase (stable for larger $\alpha$) at a threshold value 
$\alpha = {\tilde \alpha}$. 
Actually, as we are going to see, three models present infinitely 
many absorbing states.
Such fact makes standard approaches, as spreading experiments, 
difficult to use since  the dynamic exponents present values 
dependent on the initial condition \cite{dic93,munoz1,munoz3,munoz2}. 
So, in order to analyze the transition  by means
of distinct (and unambiguous) procedures, we
 first study the order-parameter $\phi$ decays 
starting from a fully occupied initial condition for
distinct independent runs. 
In the case of continuous transitions, $\phi$ decays algebraically as
$\phi \sim t^{-\theta}$ at the critical point, with $\theta$ the 
associated critical exponent. 
Conversely, at a discontinuous transition $\phi$ is not expected to 
present a power law decay. This 
 crucial difference is an important indication of the phase transition type.
To further confirm the results, we plot the probability distribution 
$P_{\phi}$ (in the steady regime) assuming different initial
configurations. 
A bimodal distribution points to a phase coexistence, whereas a single
peaked distribution -- with its position continuously
moving by changing $\alpha$ -- corresponds to a continuous transition.

\begin{figure}
\centering
\epsfig{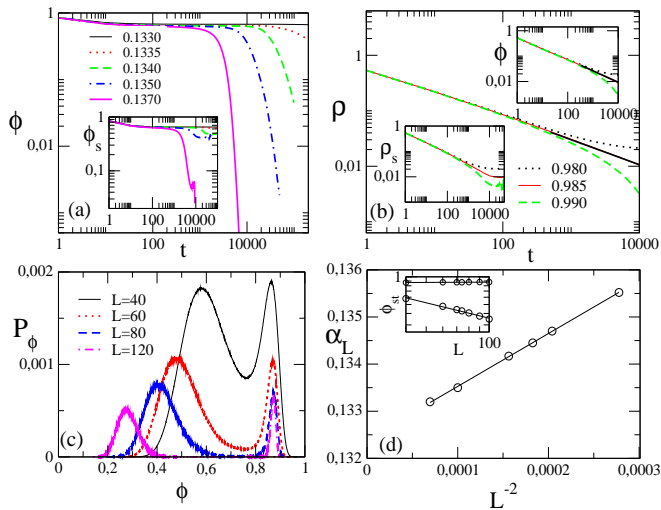}
\caption{({\bf Color online}) For the model A and  distinct $\alpha$'s, we plot in $(a)$  the time 
decay of the order parameter $\phi$   evaluated  over all   
and only survived (inset) runs, respectively. In order
to compare,  we plot in $(b)$ the time decay of the
order parameter $\rho$ for distinct $\alpha$'s by considering
the $nn \ge 1$ creation with pair annihilation \cite{adickman}.  
The black line in the middle curve has slope
$\theta=0.4505(10)$.
The upper inset in $(b)$ shows the time decay of $\phi$ (fraction 
of particles surrounded by at least $nn=1$ occupied sites) over all runs 
and the lower inset shows the time decay of $\rho$ measured
over only survived runs.   
In $(c)$ we plot the probability
distribution $P_{\phi}$ for distinct $L$'s at $\alpha_L$, in which
the peaks present the same height. The scaling plot of $\alpha_L$ 
vs $L^{-2}$ is shown in $(d)$. In the inset, we show a log-log
plot of steady order parameters $\phi_{st}$,  in the
active 
$\phi_{ac}$ and absorbing $\phi_{ab}$ phases, vs $L$. }
\label{fig1}
\end{figure}

{\it Numerical results.} Numerical simulations will be performed
in square lattices  of size $L^2$ and periodic boundary conditions.
 For the time decay analysis,
we consider $L=200$, whereas    the probability
distributions have been evaluated for $L$ ranging
from $40$ to $120$. Since isolated particles
can not created new ones, for this latter study, some extra conditions 
are required.  Following Ref. \cite{oliveira},  the extremities 
the lattice are fully occupied by particles that cannot be removed. Thus,
at any moment, there are at least four empty active sites that
providing the creation of  particles. 
Besides, since in three of four models
 isolated particles   can not create new offsprings nor be removed,
whenever the system reaches
the absorbing state a random chosen site and its nearest neighbor sites 
are fulfilled by particles.
In the first analysis,  we show in Fig. \ref{fig1} 
the main  results for the model A. 
In order to compare, we also show results
for the particle creation in the presence of $nn \ge 1$,
as studied by Dickman \cite{adickman}. A first difference 
between  the $nn \ge 1$  and $nn \ge 2$ cases 
(shown in Fig. \ref{fig1} $(b)$ and 
$(a)$, respectively) concerns that in the
latter case any configuration devoid  of pairs is absorbing and thus
the system presents infinitely many absorbing states. Unlike
the $nn \ge 1$ case, the phase transition is not ruled by   
the particle density $\rho$ for $nn \ge 2$, but for the  fraction $\phi$
of active particles (e.g. occupied sites presenting at least $nn=2$
occupied neighbors).  The second difference concerns in the time
decay behaviors. For $nn \ge 1$ (Fig. \ref{fig1}$(b)$), all curves  decay 
algebraically for low $t$,  deviating from such
behavior off the critical point for larger $t$. For $\alpha_c \sim 0.985$ 
the power law is present for sufficient large times, with an exponent 
consistent with the DP value $\theta=0.4505(10)$ \cite{marro}.  Such estimate 
for $\alpha_c$ agrees very well with the value $0.9846(1)$  
obtained by Dickman et al. \cite{adickman}. Similar exponent  
is obtained for the fraction $\phi$ of
occupied sites presenting at least $nn=1$ occupied neighbor. 
On the other hand, for $nn \ge 2$  (Fig. \ref{fig1}$(a)$) the behavior of
$\phi$ changes abruptly from a  threshold
value ${\tilde \alpha} \sim 0.1330$. For $\alpha<{\tilde \alpha}$ 
the   activity survives indefinitely,  dying off exponentially 
for $\alpha>{\tilde \alpha}$. Averages 
calculated only from survival runs enhance  above differences.
Whenever in the non restrictive case $\rho_{s}$ 
decays algebraically toward a constant value, the restriction
also provokes distinct regimes. 
For $\alpha<{\tilde \alpha}$ $\phi_{s}$  saturates
in a value close to $\phi$ (indicating
the survival of almost all runs),   whereas
for $\alpha>{\tilde \alpha}$ it decays exponentially reaching a saturated 
lower value. The  existence of discontinuous transition for $nn \ge 2$ 
is confirmed by plotting $P_{\phi}$ for distinct system sizes, as shown
in Fig. \ref{fig1}$(c)$. For all $L$'s, it 
presents a  bimodal shape, with  well defined peaks   signing
  active $\phi_{ac}$ and absorbing $\phi_{ab}$ phases. 
The former changes
very mildly with $L$, reaching the value $\sim 0.877$, 
whereas the latter vanishes following the scaling
relation $L^{-0.63(5)}$. In addition, 
the difference between  $\alpha_L$ and $\alpha_0$, in which the bimodal
probability distribution has peaks of equal height for  finite $L$ and
$L\rightarrow \infty$, respectively  scales with $L^{-2}$. 
Using this asymptotic
scale relation, we obtained the extrapolated 
value $\alpha_{0}=0.1326(2)$, which agrees
very well with the previous estimate.
We note that such dependence on the $L$ is
similar to equilibrium discontinuous  transitions \cite{rBoKo,fioreprl}.

\begin{figure}
\centering
\epsfig{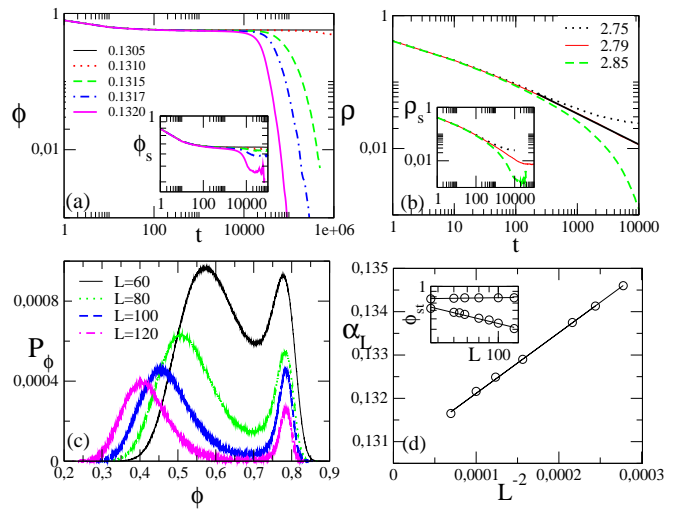}
\caption{({\bf Color online}) For the model B and  distinct $\alpha$'s, we plot in $(a)$  the time 
decay of the order parameter $\phi$   evaluated  over all   
and only survived (inset) runs, respectively. In order
to compare, we plot in $(b)$ the time decay of the
order parameter $\rho$ for distinct $\alpha$'s by considering the
$nn \ge 1$  creation case with triplet annihilation.  
The black line in the middle curve has slope
$\theta=0.4505(10)$.
The  inset in $(b)$ shows  the time decay of $\rho$ measured
over only survived runs.   
In $(c)$ we plot the probability
distribution $P_{\phi}$ for distinct $L$'s at $\alpha_L$, in which
the peaks present the same height. The scaling plot of $\alpha_L$ 
vs $L^{-2}$ is shown in $(d)$. In the inset, we show a log-log
plot of steady order parameters $\phi_{st}$,  
in the active $\phi_{ac}$ and absorbing $\phi_{ab}$ phases, vs $L$.}
\label{fig2}
\end{figure}

Next we consider the model B (exemplified
by  interaction rules $(a)-(c)$ in Fig. \ref{fig0}), 
whose results are summarized  in Fig. \ref{fig2}. 
As in the model A,  such  version    presents infinitely 
many absorbing states and the decay of the  order parameter also 
presents two distinct regimes from a threshold value 
${\tilde \alpha} \sim 0.1310$.
 For  $\alpha<{\tilde \alpha}$, it converges
to well defined value, 
indicating  indefinite activity, whereas  
the exponential decay for $\alpha>{\tilde \alpha}$ signals 
 full activity extinction. 
The pseudo transition points $\alpha_L$'s, in which the two peaks
of the probability distribution have same height,  
also scale with $L^{-2}$ for $nn \ge 2$, from
which we get the extrapolated estimate $\alpha_{0}=0.1309(1)$. Such
value agrees very well with the previous estimate ${\tilde \alpha}
\sim 0.1310$ (\ref{fig2}$(b)$).   
The dependences on $L$ of the steady order parameters
$\phi_{ac}$ and $\phi_{ab}$ are also similar than those obtained
for the previous model. Whenever $\phi_{ac}$ also changes very mildly 
with $L$, converging to the value ($\sim 0.785$), $\phi_{ac}$ vanishes
following the scaling relation $\phi_{ab} \sim L^{-0.55(8)}$, 
which is similar than the pair annihilation case.  As a result
of three particle annihilation, the compact cluster is somewhat less
compact than the value  for the model A. Despite the Elgart and
Kamenev conjecture predict  a continuous transition  (since $k=3> n=2$), numerical results show that the phase transition is first-order for
$nn\ge 2$.

\begin{figure}
\centering
\epsfig{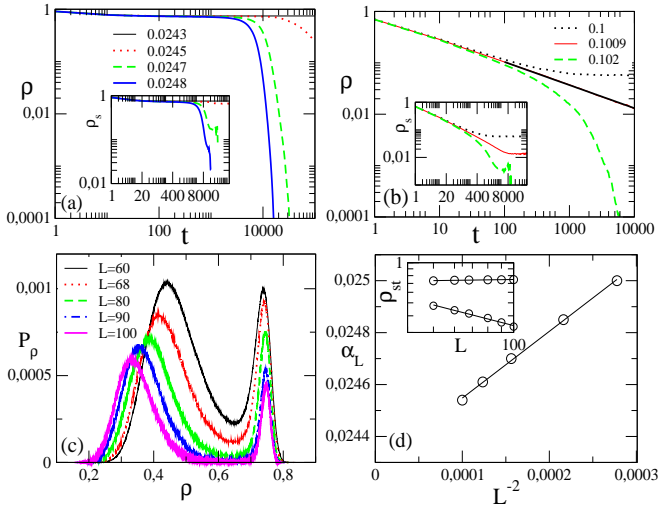}
\caption{({\bf Color online}) For the model C distinct $\alpha$'s, we plot in $(a)$  the time 
decay of the order parameter $\rho$   evaluated  over all   
and only survived (inset) runs, respectively. In order
to compare, we plot in $(b)$ the time decay of the
order parameter $\rho$ for distinct $\alpha$'s by considering the
$nn \ge 1$  creation case, but with the same annihilation rule of model C.  
The black line in the middle curve has slope
$\theta=0.4505(10)$.
The  inset in $(b)$ shows  the time decay of $\rho$ measured
over only survived runs.   
In $(c)$ we plot the probability
distribution $P_{\rho}$ for distinct $L$'s at $\alpha_L$, in which
the peaks present the same height. The scaling plot of $\alpha_L$ 
vs $L^{-2}$ is shown in $(d)$. In the inset, we show a log-log
plot of steady order parameters $\rho_{st}$,  in the active
$\rho_{ac}$ and absorbing $\rho_{ab}$ phases, vs $L$.}
\label{fig3}
\end{figure}

\begin{figure}
\centering
\epsfig{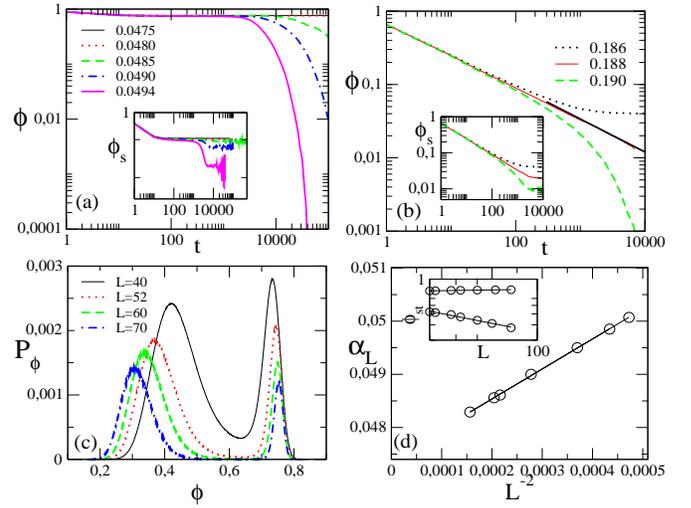}
\caption{({\bf Color online}) For the model D
and  distinct $\alpha$'s, we plot in $(a)$  the time 
decay of the order parameter $\phi$   evaluated  over all   
and only survived (inset) runs, respectively. In order
to compare, we plot in $(b)$ the time decay of the
order parameter $\phi$ for distinct $\alpha$'s for the
original PCP.
The black line in the middle curve has slope
$\theta=0.4505(10)$.
The  inset in $(b)$ shows  the time decay of $\phi$ measured
over only survived runs.   
In $(c)$ we plot the probability
distribution $P_{\phi}$ for distinct $L$'s at $\alpha_L$, in which
the peaks present the same height. The scaling plot of $\alpha_L$ 
vs $L^{-2}$ is shown in $(d)$. In the inset, we show a log-log
plot of steady order parameters $\phi_{st}$,  in the active
$\phi_{ac}$ and absorbing $\phi_{ab}$ phases, vs $L$.
 }
\label{fig4}
\end{figure}

In order to strengthen the above conclusions, we examine the model C, 
whose extinction includes all neighboring occupied
sites of a given particle chosen at random.   
Unlike the previous examples,
the system presents a single absorbing state and thus the dynamics is
ruled by the particle density $\rho$. The results are summarized in
Fig. \ref{fig3}.
As in the previous examples, the creation in the presence
of $nn \ge 1$ (Fig. \ref{fig3} $(b)$) and $nn\ge 2$ (Fig. \ref{fig3} $(a)$) 
behave very differently.
Whenever in the former, $\rho$  decays following a
 DP  exponent $\theta=0.4505(10)$
at $\alpha_c \sim 0.1009$, for $nn\ge 2$ 
one has   two distinct regimes separated 
from a given threshold value ${\tilde \alpha} \sim 0.0244$.  
The probability distribution $P_{\rho}$ (Fig. \ref{fig4} 
$(c)$) is also bimodal for $nn \ge 2$, and
the positions of two equal peaks $\alpha_L$'s also scale 
with $L^{-2}$, from which one gets the estimate $\alpha_0=0.0243(1)$ - in 
excellent agreement with ${\tilde \alpha}$. 
The   steady densities $\rho_{st}$'s also  
exhibit distinct dependences on the system size and are similar 
than previous cases. Whenever $\rho_{ac}$  
saturates in a constant value $\rho_{ac}\sim 0.747$ 
when $L$ increases,   $\rho_{ab}$ vanishes according
to the asymptotic law  $L^{-0.52(5)}$.

Last, we extend the restriction for the two-dimensional
PCP (model D).  
By comparing Figs. \ref{fig4} $(a)$ and $(b)$, we see that in similarity
with previous models, the order parameter $\phi$ (the pair density 
in both cases) behaves
differently in the original $nn_{p} \ge 1$ and $nn_{p}\ge 2$ versions. 
Whenever in the
former $\phi$ decays with an exponent consistent with the DP value
$\theta=0.4505(10)$ at the phase transition (placed at $\alpha_c \sim
0.188$ \cite{note1}), a threshold value (${\tilde \alpha}\sim 0.0480$)
separates permanent ($\alpha<{\tilde \alpha}$) from the full activity
extinction ($\alpha >{\tilde \alpha}$) for $nn_{p}\ge 2$.  Averages evaluated
over survival runs corroborate the differences between both versions
as well as the similarities among above three examples.  The
probability distribution $P_{\phi}$ (Figs. \ref{fig4} $(c)$) also
presents two peaks for $nn_{p} \ge 2$, whose $\alpha_L$'s scale with $L^{-2}$, 
providing the extrapolated estimate $\alpha_0=0.0474(2)$, which is close
to the above estimate.  As in all previous restrictive examples,
$\phi_{ac}$ and $\phi_{ab}$ also exhibit distinct dependences on the
system size $L$.  Whenever $\phi_{ac}$ reaches the constant value
$\sim 0.755$ in the thermodynamic limit, $\phi_{ab}$ vanishes
according to the relation $L^{-0.63(5)}$, in consistency with all
previous examples.

{\it Conclusion.} 
To sum up, we presented strong evidences of a minimal mechanism
 leading to a first-order transition into absorbing states for short
range systems.
In all cases, results differing greatly from their original  cases 
(in which the phase transitions are unambiguously  continuous) have
been achieved.
The onset of a threshold value  separating   
endless activity from  an exponential decay toward the full extinction 
as well as   bimodal distributions with (pseudo-) transition points scaling
on the system volume  strongly suggests that 
the particle creation 
in the presence of a minimal neighborhood (for all studied models it is 2) 
constitute a robust and fundamental ingredient
determining the phase coexistence in short-range contact processes.
An understanding about the role of such particle creation
requirement is achieved by performing mean field calculations.
By taking correlation at level of two sites, 
in all cases   low density states become unstable 
for low $\rho$ (with $\alpha$
increasing with $\rho$), signaling  a  jump. On the other hand,
for the non restrictive versions, $\rho$ always
decreases with $\alpha$. 
As a final remark, we note that the study
of other  restrictive processes, including the diffusion
of particles 
and  competitive dynamics should be addressed
in a previous contribution.

{\bf Acknowledgment.} I acknowledge 
Gandhi Viswanathan, M. W. Beims and M. G. E. da Luz for  critical readings 
of this manuscript and  the research  grant from  CNPQ.


\begin{thebibliography}{99}
\bibitem{note1}  In fact, 
our estimate is   different to the estimate $\alpha_c=0.20051(1)$ 
\cite{dic99},
 as a result of a slightly different dynamics implementation.
\end{thebibliography}
\end{document}